\documentclass[aps,prb,preprint,superscriptaddress,colorlinks=true,linkcolor=black,urlcolor=black,citecolor=black,anchorcolor=black,colorlinks,bookmarks=false,citecolor=blue,linkcolor=black,urlcolor=blue]{revtex4-1}

\usepackage{ulem}
\usepackage{bm,color,amsmath,amssymb,latexsym,graphicx,psfrag,accents,float}
\usepackage{amscd,dsfont,wasysym,mathrsfs,mathtools}
\usepackage{multirow}
\usepackage{dcolumn}
\usepackage{xcolor}
\usepackage{comment}
\usepackage{booktabs} 
\usepackage{upgreek}
\usepackage[utf8]{inputenc}
\usepackage{braket}
\usepackage{times}
\usepackage{amsfonts}
\usepackage{mathrsfs}
\usepackage{graphicx}
\usepackage{dcolumn}
\usepackage{bm}
\usepackage{color}

\usepackage[colorlinks,bookmarks=false,citecolor=blue,linkcolor=red,urlcolor=blue]{hyperref}
\usepackage{amsmath}
\usepackage{ulem}
\begin{document}
\renewcommand{\theequation}{S\arabic{equation}}
	\newcommand{\beginMethods}{%
		\setcounter{table}{0}
		\renewcommand{\thetable}{S\arabic{table}}%
		\setcounter{figure}{0}
		\renewcommand{\thefigure}{S\arabic{figure}}%
	}
\newcommand{\bk}{{\bf k}}
\newcommand{\bB}{{\bf B}}
\newcommand{\bv}{{\bf v}}

\title{Correlated order at the tipping point in the kagome metal CsV$_3$Sb$_5$}

\author{Chunyu Guo${}^{\dagger}$}\affiliation{Max Planck Institute for the Structure and Dynamics of Matter, Hamburg, Germany}
\author{Glenn Wagner}\affiliation{Department of Physics, University of Zürich, Zürich, Switzerland}
\author{Carsten Putzke${}^{}$}
\affiliation{Max Planck Institute for the Structure and Dynamics of Matter, Hamburg, Germany}
\author{Dong Chen}\affiliation{Max Planck Institute for Chemical Physics of Solids, Dresden, Germany}\affiliation{College of Physics, Qingdao University, Qingdao, China}
\author{Kaize Wang${}^{}$}
\affiliation{Max Planck Institute for the Structure and Dynamics of Matter, Hamburg, Germany}
\author{Ling Zhang${}^{}$}
\affiliation{Max Planck Institute for the Structure and Dynamics of Matter, Hamburg, Germany}
\author{Martin Gutierrez-Amigo}\affiliation{Centro de Física de Materiales (CSIC-UPV/EHU), Donostia-San Sebastian, Spain}
\affiliation{Department of Physics, University of the Basque Country (UPV/EHU), Bilbao, Spain}
\author{Ion Errea}\affiliation{Centro de Física de Materiales (CSIC-UPV/EHU), Donostia-San Sebastian, Spain}
\affiliation{Donostia International Physics Center, Donostia-San Sebastian, Spain}
\affiliation{Fisika Aplikatua Saila, Gipuzkoako Ingeniaritza Eskola, University of the Basque Country (UPV/EHU), Donostia-San Sebastian, Spain}
\author{Maia G. Vergniory}\affiliation{Donostia International Physics Center, Donostia-San Sebastian, Spain}
\affiliation{Max Planck Institute for Chemical Physics of Solids, Dresden, Germany}
\author{Claudia Felser}\affiliation{Max Planck Institute for Chemical Physics of Solids, Dresden, Germany}
\author{Mark H. Fischer${}^{\dagger}$}\affiliation{Department of Physics, University of Zürich, Zürich, Switzerland}
\author{Titus Neupert${}^{\dagger}$}\affiliation{Department of Physics, University of Zürich, Zürich, Switzerland}
\author{Philip J. W. Moll${}^{\dagger}$}\affiliation{Max Planck Institute for the Structure and Dynamics of Matter, Hamburg, Germany}

\date{\today}
\maketitle
\normalsize{$^\dagger$Corresponding authors: chunyu.guo@mpsd.mpg.de(C.G.); mark.fischer@uzh.ch(M.H.F.); titus.neupert@uzh.ch(T.N.); philip.moll@mpsd.mpg.de(P.J.W.M.).}

\section*{ABSTRACT}
Spontaneously broken symmetries are at the heart of many phenomena of quantum matter and physics more generally. 
However, determining the exact symmetries broken can be challenging due to imperfections such as strain, in particular when multiple electronic orders form complex interactions. 
This is exemplified by charge order in some kagome systems, which are speculated to show nematicity and flux order from orbital currents.
We fabricated highly symmetric samples of a member of this family, CsV$_3$Sb$_5$, and measured their transport properties. We find the absence of measurable anisotropy at any temperature in the unperturbed material, however, a striking in-plane transport anisotropy appears when either weak magnetic fields or strains are present.
A symmetry analysis indicates that a \textit{perpendicular} magnetic field can indeed lead to \textit{in-plane} anisotropy by inducing a flux order coexisting with more conventional bond order. 
Our results provide a unifying picture for the controversial charge order in kagome metals and highlight the need for microscopic materials control in the identification of broken symmetries.

\section*{MAIN TEXT}
\subsection*{Introduction}
Materials hosting intertwined electronic ordering phenomena provide both an outstanding challenge to and opportunity in current condensed matter physics. Phases such as magnetism, charge order, spin textures or superconductivity may cooperate or compete, or merely coexist without much commonality, thus complicating interpretation of experimental data\cite{HFRMP,cupratesRMP}. Disentangling the various order parameters into the most elemental building blocks, separating the primary from secondary orders, and understanding their interrelation, is the key to solving their rich puzzle. Complex quantum materials with entangled correlated phases also offer unique electronic response functions: When manipulating one order, one can switch or tune a different one, akin to the unique electromagnetic responses of multi-ferroics\cite{tokura2014multiferroics,Cheong2007multiferroics}.

With the poor performance of ab-initio predictions for such correlated materials, a promising experimental route is to follow structural motifs that either host highly degenerate states or exhibit geometrically frustrated bonds, which are known to commonly host correlated phases at low temperatures. Honeycomb lattices, square nets, perovskite cages, or the kagome lattice have been extremely fruitful in this regard~\cite{KagomeReview,FrustrationReview,Cheong2007multiferroics,Keimer2017QM,SpiniceRMP}. The kagome lattice, a net of triangles connected at their vertices, combines both bond frustration and sublattice symmetries and thus, has been a successful platform for the design of nontrivial quantum materials\cite{syozi1951statistics,kang2020topological,kang2020,Yin2020,Yin2018,Ye2018,Howard2021,ortiz2019new}.

Recently, the kagome family (Cs,K,Rb)V$_3$Sb$_5$ has received significant attention due to the wealth of phases it hosts\cite{ortiz2019new,ortiz2020cs,Kang2022,TitusAdd,Neupert2022}. At high temperature, these V-based kagome systems are in a centrosymmetric, non-magnetic metallic state, but undergo a charge-density-wave (CDW) type instability at $T_{\rm CDW}\sim 100$K. Consensus of a 2x2 reconstruction in the kagome plane has been reached, yet the exact low-temperature structure and the nature of the out-of-plane reconstruction (2x2x2 vs 2x2x4) remain to be clarified\cite{zhao2021cascade,chen2021roton,CVS_Xray}. At even lower temperature, superconductivity appears at a critical temperature $T_{\rm c}$ which is enhanced by hydrostatic pressure, anti-correlated with the pressure dependence of the CDW indicating their competitive relationship\cite{PressureAM,du2021pressure,Zheng2022}. Beyond this, an impressive set of experiments has demonstrated that ``something else'' is happening in this material, often associated with an onset temperature of $T'\sim 20-50$K, yet the nature and physical influences are under heavy debate, fueled by openly contradictory experimental results. These experiments include signatures of time-reversal-symmetry (TRS) breaking~\cite{Mielke2022,yu2021evidence,Kerr_TRSyes,Xu2022} and absence thereof\cite{TRSno}; electronic nematicity\cite{xiang2021twofold,Nie2022,Nema2,NLWang_twofold}; tunable chirality\cite{Kchiral,Guo2022} and absence thereof\cite{Li2022}.

The central question is how such carefully conducted experiments on a deceptively simple, stoichiometric material of high crystalline purity yield such contradictory results. Here, we propose, and experimentally demonstrate, that these discrepancies are intrinsically rooted in the strong coupling of the multitude of orders it hosts. This renders these material class extraordinarily sensitive to even weak perturbations, which could safely be ignored in conventional compounds. We demonstrate that in-plane strain and magnetic fields are such sensitive perturbations, yet the material may be highly sensitive to others as well. In the experimental reality, weakest residual strains (e.g., from crystal defects and sample mounting) are ubiquitous and hard to avoid, while magnetic fields of several Tesla are used in the spirit of non-invasive probes (quantum oscillations, NMR, magnetotransport). We propose that moving towards maximally decoupled crystalline samples will consolidate the field and lead towards the identification of the correlated ground state without perturbations. The main message of this paper is that the at first glance contradictory state of the literature is a feature, not a bug. (Cs,K,Rb)V$_3$Sb$_5$ may well realize the long-standing dream of exotic electronic response functions that directly arise from the near-degeneracy of multiple distinct correlated states.

\subsection*{Isotropic in-plane transport in zero field}

One proposed electronic instability is a spontaneous breaking of the six-fold rotational symmetry of the kagome plane into a two-fold symmetric electronic state\cite{NLWang_twofold,xiang2021twofold,Nie2022}. Such nematic transitions manifest in an emergent transport anisotropy within the plane, which is symmetry-forbidden in a six-fold symmetric kagome plane. To this end, we have machined hexagon-shaped microstructures featuring six contacts, one at each corner of the hexagon, from bulk single crystals using Focused-ion-beam (FIB) milling. Great care was taken to align the structure with the in-plane lattice vectors via XRD ($<$ 0.5 deg) and to minimize shape deviations to avoid any symmetry lowering due to the structure's shape itself (Fig. 1). Systematic resistance measurements were performed with current applied diagonally across the hexagon and voltage measured along the side, in all 3 possible configurations. As we will show even weak strains to be critical factors, it is important to remove any residual strain onto the micro-shaped crystal that arises from differential thermal contraction. 

Low-strain samples have been achieved in two ways, either by decoupling the structure from the substrate by suspending it freely on ultra-soft SiN membranes (Device S1); or by encasing it in an epoxy droplet crafted in such a way that the compressive forces of the contracting soft glue compensate the tensile forces of a hard substrate coupling (Device S3). Both low-stress designs result in an approximately strain-free hexagon at cryogenic temperatures (See supplement for residual strain estimates and fabrication details). To demonstrate the role of directional strain, a sample on a purposely mismatched substrate was subjected to unidirectional strain due to their differential thermal contraction (Device S2) \cite{Maja,Maarten}. Sharp transitions at $T_{\rm CDW}$ and $T_{\rm c}$, in perfect agreement with the bulk values, signal the approximately strain-free state of S1 and S3, while our strained device S2 differs in $T_{\rm CDW}$ and $T_{\rm c}$ according to their known strain dependence (see supplement). Our set of samples allows us to probe microscopic signatures of rotational symmetry breaking with a high level of control over the physical state of the sample.



For temperatures above $T_{\rm CDW}$, the resistances measured along three current directions are identical in each of the three samples (Fig. 2), as expected for a six-fold symmetric kagome metal. Interestingly, they remain identical for the low-strain samples S1 and S3 at all temperatures down to $T_{\rm c}$within $\pm$ 0.05\%. In light of the various contradictory results on the spontaneous symmetry breaking of the electronic structure in CsV$_3$Sb$_5$, we emphasize the argumentative power of observing the absence of anisotropy. Electronic anisotropy, if observed, may appear due to intrinsic or extrinsic symmetry breaking, yet its absence constrains us to two possible microscopic scenarios of the underlying electronic state: Either the sample decays into perfectly balanced domain structures as to restore the apparent C6 symmetry of macroscopically averaging probes such as transport; or it remains fundamentally not broken in the absence of perturbations. The reproducibility of these results in combination with the 10 $\mu m$ size of the hexagon necessitates extremely small domains in the nm-range, as to obtain reliably perfect averaging. Both the high density of energetically costly domain boundaries and the absence of such vastly textured electronic matter in local-probe experiments\cite{Xu2022,zhao2021cascade} speak against the nano-domain picture, while we note it cannot be excluded from our data. 

A different picture appears in the strained case: At $T_{\rm CDW}$, a clearly observable resistivity anisotropy immediately appears. Two directions remain identical within experimental accuracy, while the one along the unidirectional strain direction differs. We quantify the experimental anisotropy as the anti-symmetric difference between these two directions, $(R_a - R_b)/(R_a + R_b)$. This anisotropy remains relatively featureless until it distinctly grows around $T'\sim 30$ K to above 30\% at $T = 5$ K. This sudden growth in the strained device S2 is accompanied by a sign change of the anisotropy, occurring simultaneously with the various anomalies reported around $T'$. 


\subsection*{Field- and Uniaxial-strain-induced anisotropy}


A second, key piece to the puzzle is unveiled by magnetic fields. Previous reports\cite{Kchiral} suggested that the rotational symmetry breaking is linked to time-reversal-symmetry breaking. Under a static out-of-plane magnetic field $B_c$, even strain-free samples exhibit a transport anisotropy, which increases monotonically with increasing magnetic field (Extended Data Fig. 4). For $B_c = 9$~T, the anisotropy onsets at $T~\approx$ 70 K and reaches 20$\%$ at low temperature. Similar to the strained samples, the anisotropy increases markedly around $T' \approx$ 35 K. Detailed rotation studies of the magnetic field away from the $c$ direction have shown no influence, hence we can experimentally exclude a small accidental in-plane field component to be the source of the symmetry breaking (Extended Data Fig. 7). Repeated cooldowns exactly reproduce the state of the sample, speaking against spontaneous symmetry breaking and likely the very weak but non-zero residual strain breaks the symmetry. Still, in the same sample without a magnetic field no anisotropy is detectable. This behavior marks out-of-plane magnetic fields $B_c$ as another crucial tuning parameter for electronic symmetry breaking in CsV$_3$Sb$_5$. Interestingly, the nematic state can be directly controlled by a magnetic field, evidencing the strong coupling between these phenomena. Together, these observations substantiate a consistent experimental picture of a $C_6$ symmetric state in the pristine material that is critically unstable against forming a nematic state under weak perturbations. Indeed, when strain and out-of-plane fields are simultaneously applied, the anisotropy grows even further (Extended Data Fig. 6).


\subsection*{Theoretical model of symmetry analysis and strain-field coupling}
While uniaxial in-plane strain can naturally produce  an anisotropy by explicitly breaking rotation symmetry, the effect of an out-of-plane field on the transport anisotropy appears at first glance surprising. In the following, we elucidate the coupling of strain and magnetic field on a charge-density order parameter in kagome systems within Ginzburg-Landau (GL) theory.
For this discussion, we focus on a single kagome layer and are guided by the following experimental observations: (1) There is translational symmetry breaking in the form of a $2\times2$ increase of the size of the (in-plane) unit cell; (2) There is no spontaneous rotational symmetry breaking in the absence of any external perturbation. This restriction will fix the terms of the GL expansion to fourth order in powers of the order parameter; And (3) the magnetic field couples linearly to the system, which restricts the form of the flux phase;
We introduce a complex three-component order parameter $\vec{\psi}=\vec{\Delta}+i\vec{\Delta}'$ defined in Fig.~2(a) (and transformation properties defined in supplementary table I). $\vec{\Delta}$ describes charge bond order while $\vec{\Delta}'$ describes flux order and therefore breaks time-reversal symmetry. Importantly, this flux order is even under $C_2$ symmetry, which is crucial for observation (3). The resulting GL free energy, shown in the supplement, is extremely rich and includes not only third-order coupling \cite{Christensen,Denner,Grandi}, but also a linear coupling of $\vec{\Delta}$ with $\vec{\Delta}'$ mediated by the out-of-plane field~\cite{Tazai}. In the absence of a magnetic field and strain, this free energy leads to the phase diagram shown in Fig.~2(b), where from left to right we tune the relative critical temperature of the bond and flux order. As shown in the cuts in Fig.~2(c), due to the presence of a third order term in the free energy, any $\vec{\Delta}'$ induces a subsidiary $\vec{\Delta}$, while the converse is not true. Finally, the third-order terms coupling the two order parameters results in an anisotropic solution. Given our observation (2), namely an isotropic order in the absence of external perturbations, we will focus in the following on the scenario 4 of the phase diagram presented in Fig.~2(b).


\subsection*{Field dependence of anisotropy and phase diagram}

The above symmetry considerations on the Ginzburg-Landau level provide a field-strain-temperature phase diagram which reproduces the experimental observations remarkably well. The bond order $\vec{\Delta}$ appears below $T_\textrm{CDW}$ at a (weakly) first-order transition due to the presence of the third-order term. In the absence of a magnetic field or strain, there is no anisotropy, in other words the three components of $\vec{\Delta}$  are identical. When a magnetic field is applied, $\vec{\Delta}'$ is induced leading to a non-zero anisotropy due to the third-order coupling of $\vec{\Delta}$ and $\vec{\Delta}'$. This anisotropy becomes observable at a temperature below the charge-ordering temperature, when the third-order term becomes large enough. The anisotropy is further enhanced by increasing the magnetic field, since that increases the $\vec{\Delta}'$ component. In the presence of strain, an anisotropy in $\vec{\Delta}$ is induced immediately at the charge ordering temperature. Furthermore, at a lower temperature $\vec{\Delta}'$ can condense as well in the presence of strain, which increases the anisotropy. In both cases, TRS breaking only occurs when $\vec{\Delta}'$ is present. 

In our scenario, pristine CsV$_3$Sb$_5$ is thus time-reversal symmetric at any temperature, yet located critically close to the TRS broken phase $\vec{\Delta}+i\vec{\Delta}'$ in the phase diagram (Fig. 3). Applying the magnetic field directly promotes the coupling between the orders and drives an in-plane symmetry breaking at arbitrary low fields, i.e.\ without a threshold field. This scenario indeed matches the experimental data:
Starting at zero in the absence of field or strain, the anisotropy immediately appears under field and continuously increases up to $B$ = 2~T. At higher magnetic field, the anisotropy continues to grow yet at a slower pace and exhibits pronounced quantum oscillations. While microscopic details such as Landau quantization are naturally not considered in a GL theory, it is remarkable that its general trend continues despite the highly complex transport situation in the microstructure.


Finally, we note that raising the temperature pushes the system towards an isotropic state, which can be straightforwardly mapped out in our experiment (Fig. 3b). The thin isotropic sliver immediately gives rise to an anisotropic state under weakest magnetic fields, which is remarkably similar to the predictions of the GL theory shown in Fig.~3c. This behavior can be directly rationalized from the phase diagram in Fig.~2b. In such a picture, raising the temperature pushes the system deeper into the isotropic, time-reversal-symmetric state.



\subsection*{Discussion and Outlook}

There have been numerous reports on the spontaneous symmetry breaking in CsV$_3$Sb$_5$ (Fig. 4). Despite the variety of experimental methods, most of the results recognize a six-fold to two-fold in-plane symmetry lowering upon entering the charge-ordered state. In contrast, our measurements of the nearly-strain-free samples clearly demonstrate that the electronic transport preserves six-fold symmetry across the whole temperature range. We argue that this apparent discrepancy is mainly due to the fact that in most experiments, such as angular dependence of magnetoresistance and elasto-transport measurements, magnetic fields and uniaxial strain act as necessary probes of symmetry breaking/lowering. The extreme sensitivity to perturbations amplifies even accidental and in-adverted perturbations that may arise from crystal growth, handling or mounting in the experimental reality. Here, we have clearly demonstrated that these presumably small perturbations act as crucial tuning parameters of the transport anisotropy (Fig. 4). Importantly, there exist two critical temperature scales. Except for the apparent transition temperature $T_{\rm CDW}$ at which the strain-induced anisotropy appears exactly, the additional temperature scale $T$' stands for the onset of the field-induced anisotropy. This is consistently demonstrated not only by our measurements but also all experiments with external magnetic field or uniaxial strain. These two distinct temperature scales suggest that despite the similar effect between strain and field, the way they couple to the order is different. 

Our Ginzburg-Landau discussion provides an intriguing scenario that may reconcile some of the apparent contradictions reported in this field. In this picture, CsV$_3$Sb$_5$ exhibits an extreme sensitivity of the transport anisotropy due to its critical proximity to the loop-current phase. The anisotropy in the order parameters for cut 4 through the phase diagram in Fig.~2b reproduces successfully the influence of a magnetic field and strain on the temperature dependence of the anisotropy. Hence, our experiments, as well as the published literature, are consistent with a regime where the time-reversal-preserving bond order is dominant, while the TRS-breaking flux order is subdominant and is only induced in the perturbed case. 

Our work once more highlights the rich responses and possible advanced functionalities in materials hosting a subtle balance of entangled orders, which can mutually be manipulated. From a larger perspective, this work sends a beacon of hope into controversial fields which similarly may be unified when previously uncontrolled tuning parameters become controllable in the future, no matter how negligible and irrelevant they may seem at first glance.

%

\clearpage

\begin{figure}
	\centering
        \includegraphics[width = 0.77\linewidth]{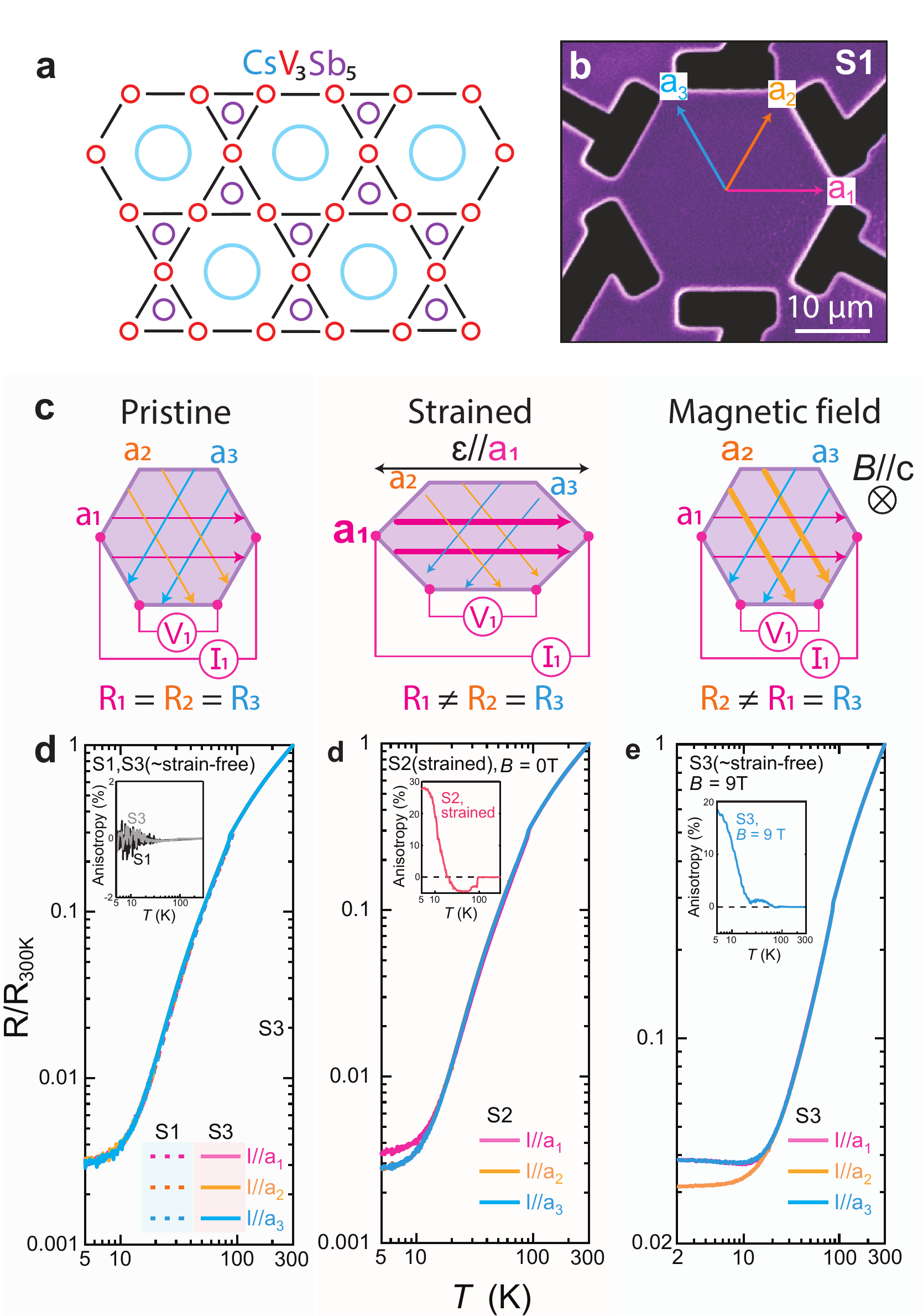}
		\caption{\textbf{Field- and strain-induced in-plane transport anisotropy in CsV$_3$Sb$_5$.} (a) Detailed setup of tri-directional resistance measurement and possible origin of in-plane anisotropy. (b) Scanning electron microscopy (SEM) image of device S1. The hexagon of CsV$_3$Sb$_5$ is fabricated via focused-ion-beam technique with six symmetric contacts. Temperature dependence of tri-directional resistance for (c) the nearly strain-free devices S1 and S3 in zero field, (d) device S2 with ineligible uniaxial strain and (e) device S3 in the presence of magnetic field ($B$ = 9 T).}
	\label{Intro}
\end{figure}
\clearpage
\begin{figure}
	\centering
        \includegraphics[width = 0.9\linewidth]{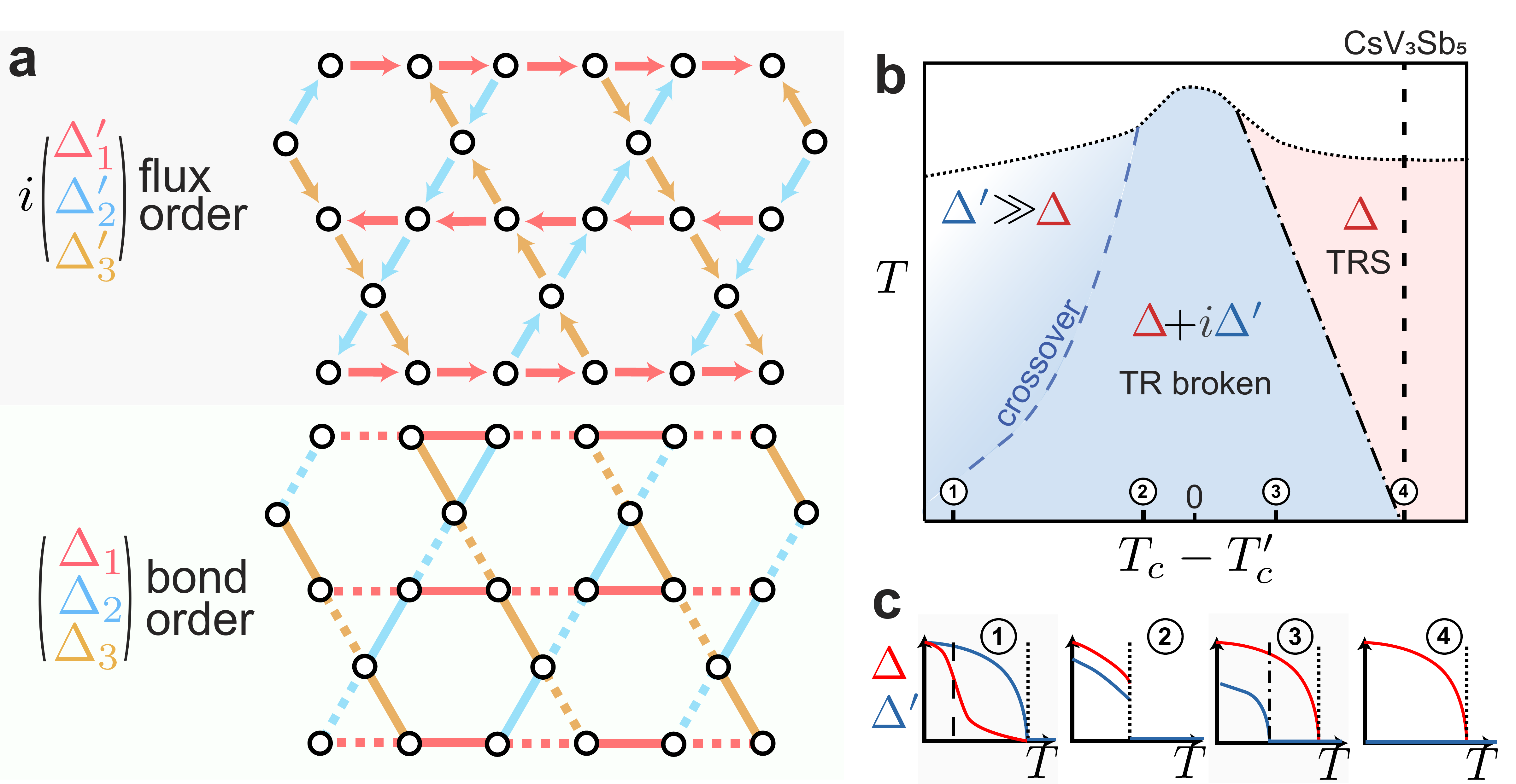}
		\caption{\textbf{Ginzburg-Landau phase diagram.} (a) Definition of the three-component order parameters $\vec{\Delta}$ (describing real modulation of the bond hopping) and $\vec{\Delta}'$ (describing imaginary modulation of the bond hopping, meaning flux order). Both order parameters break translational symmetry and increase the unit cell by $2\times 2$. In the lower panel, solid and dashed lines denote positive and negative values of the bond order parameter, respectively. (b) Schematic phase diagram of the order parameters $\vec{\Delta}$ and $\vec{\Delta}'$. From left to right, we tune the relative critical temperature of $\vec{\Delta}$ and $\vec{\Delta}'$. Time-reversal symmetry (TRS) is broken in the blue region, whereas we have TRS in the pink region. (c) Order parameters along four vertical cuts through the phase diagram. $\vec{\Delta}'$ always induces a subsidiary $\vec{\Delta}$, while the converse is not true.}
	\label{Cs135}
\end{figure}
\clearpage
\begin{figure}
	\centering
        \includegraphics[width = 0.95\linewidth]{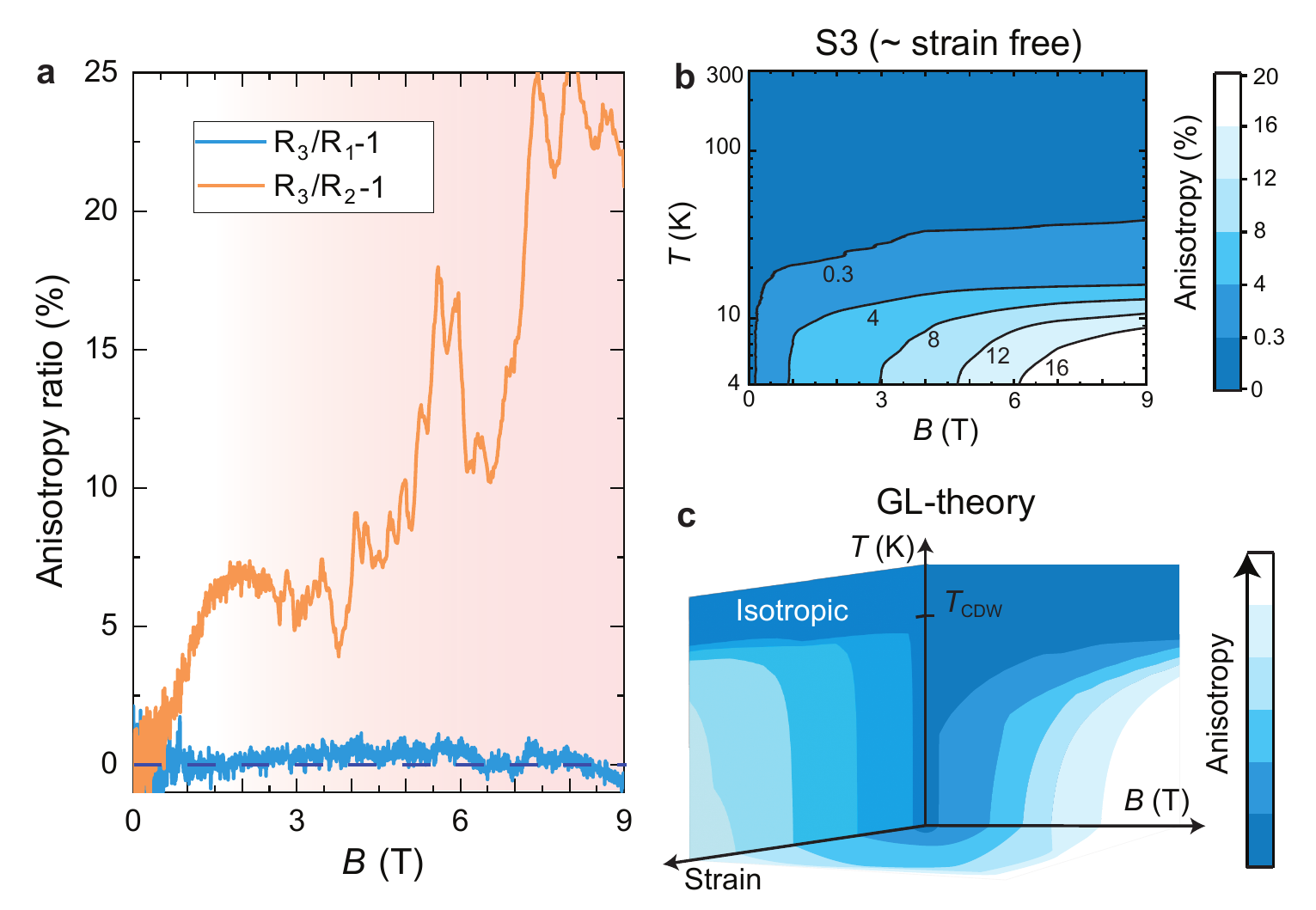}
		\caption{\textbf{Anisotropic magnetoresistance and characteristic magnetic field scale.} (a) Field dependence of transport anisotropy in CsV$_3$Sb$_5$. (b) Temperature-field phase diagram of anisotropy. The anisotropy stays zero without applying external magnetic field ($B$ = 0 T) and above $T$' = 40 K. (c) Temperature-field-strain phase diagram reproduced based on Ginzburg-Landau theory, which is consistent with the experimental results displayed in (b).}
	\label{Theory}
\end{figure}
\clearpage
\begin{figure}
	\centering
        \includegraphics[width = 0.55\linewidth]{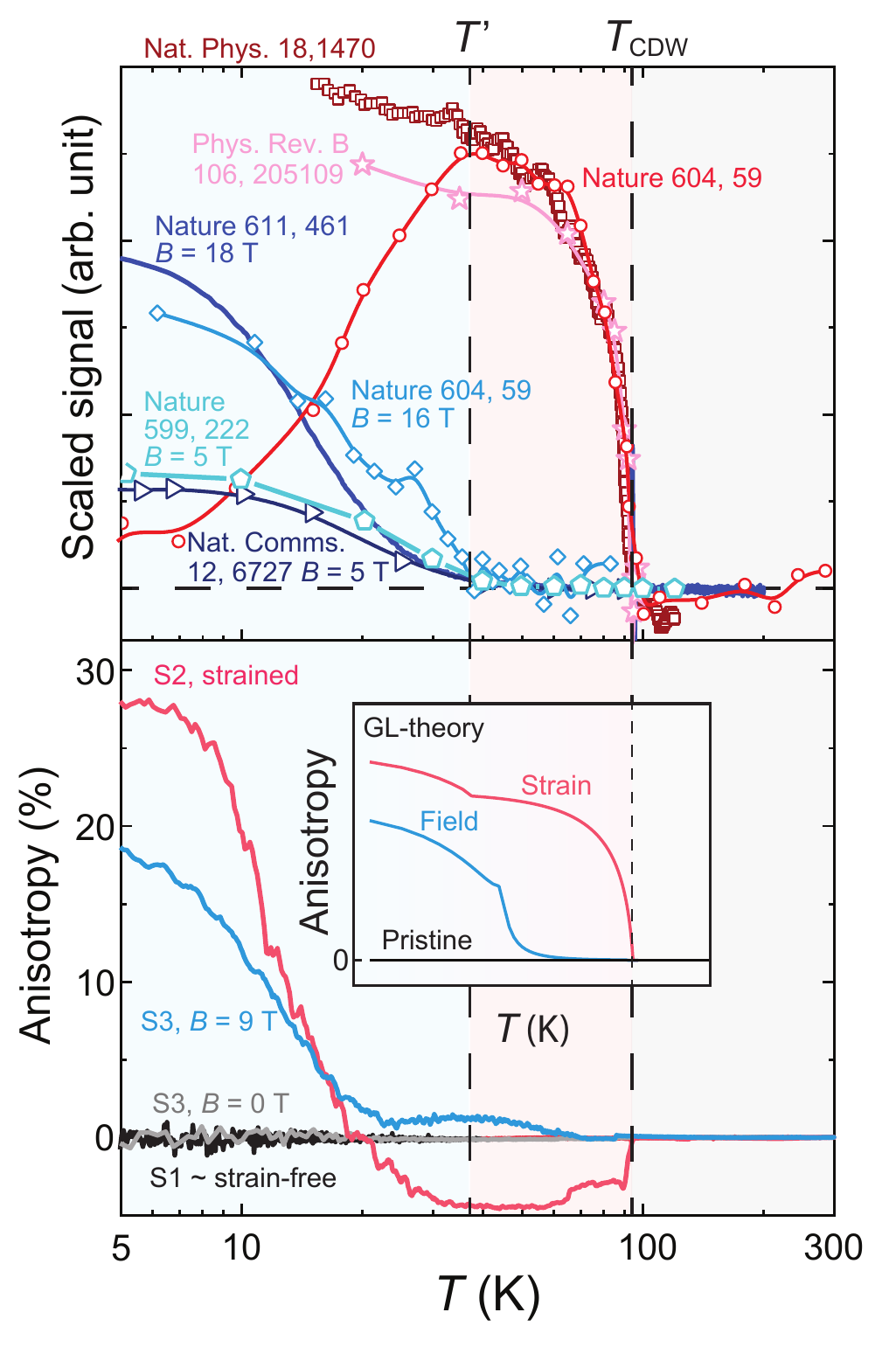}
		\caption{\textbf{Summary and comparison with previous reports.} The upper panel summarizes the previously reported temperature dependence of anisotropy measured via various methods, including angle-dependent magnetoresistance\cite{chen2021roton,xiang2021twofold}, electric magneto-chiral anisotropy\cite{Guo2022}, elasto-resistivity\cite{Nie2022}, NMR\cite{Zheng2022}, magneto-optical Kerr effect\cite{Xu2022} and optical polarization rotation measurement\cite{NLWang_twofold}. Interestingly, all results measured without magnetic field demonstrate that the anisotropy occurs once the charge-order is formed. Meanwhile the in-field measurements consistently reveal yet another temperature scale ($T$') below $T_{CDW}$. The lower panel displays the anisotropy measured in device S1, S2 and S3. The field-induced anisotropy occurs only below $T_{CDW}$ and displays a broad hump centered at $T$', while the strain-induced anisotropy onsets exactly at $T_{CDW}$. The inset panel presents the T-dependent anisotropy of the order parameter under different conditions reproduced by Ginzburg-Landau theory. We choose Ginzburg-Landau parameters corresponding to scenario \textcircled{4} in Fig. 2(b)(c).}
	\label{Main}
\end{figure}
\clearpage

\noindent \textbf{Acknowledgements: } This work was funded by the European Research Council (ERC) under the European Union’s Horizon 2020 research and innovation programme (MiTopMat - grant agreement No. 715730 and PARATOP - grant agreement No. 757867). This project received funding by the Swiss National Science Foundation (Grants  No. PP00P2\_176789). M.G.V., I. E. and M.G.A. acknowledge the Spanish   Ministerio de Ciencia e Innovacion  (grant PID2019-109905GB-C21). M.G.V., C.F., and T.N. acknowledge support from FOR 5249 (QUAST) lead by the Deutsche Forschungsgemeinschaft (DFG, German Research Foundation). M.G.V. acknowledges partial support to  European Research Council grant agreement no. 101020833. This work has been supported in part by Basque Government grant IT979-16. This work was also supported by the European Research Council Advanced Grant (No. 742068) “TOPMAT”, the Deutsche Forschungsgemeinschaft (Project-ID No. 247310070) “SFB 1143”, and the DFG through the W\"{u}rzburg-Dresden Cluster of Excellence on Complexity and Topology in Quantum Matter ct.qmat (EXC 2147, Project-ID No. 390858490).

\noindent \textbf{Author Contributions:} Crystals were synthesized and characterized by D.C. and C.F.. The experiment design, FIB microstructuring, the magnetotransport measurements were performed by C.G., C.P., S.K., X.H. and P.J.W.M.. G.W., M.H.F. and T.N. developed and applied the general theoretical framework, and the analysis of experimental results has been done by C.G., C.P. and P.J.W.M.. All authors were involved in writing the paper.

\noindent \textbf{Competing Interests:} The authors declare that they have no competing financial interests.\\

\noindent \textbf{Data Availability:} Data that support the findings of this study will be deposited to Zenodo with the access link displayed here.

\clearpage

\end{document}


\renewcommand{\theequation}{S\arabic{equation}}
	\newcommand{\beginsupplement}{%
		\setcounter{table}{0}
		\renewcommand{\thetable}{S\arabic{table}}%
		\setcounter{figure}{0}
		\renewcommand{\thefigure}{S\arabic{figure}}%
	}
\title{Supplementary materials for "Correlated order at the tipping point in the kagome metal CsV$_3$Sb$_5$"}

\author{Chunyu Guo${}^{\dagger}$}\affiliation{Max Planck Institute for the Structure and Dynamics of Matter, Hamburg, Germany}
\author{Glenn Wagner}\affiliation{Department of Physics, University of Zürich, Zürich, Switzerland}
\author{Carsten Putzke${}^{}$}
\affiliation{Max Planck Institute for the Structure and Dynamics of Matter, Hamburg, Germany}
\author{Kaize Wang${}^{}$}
\affiliation{Max Planck Institute for the Structure and Dynamics of Matter, Hamburg, Germany}
\author{Ling Zhang${}^{}$}
\affiliation{Max Planck Institute for the Structure and Dynamics of Matter, Hamburg, Germany}
\author{Martin Gutierrez-Amigo}\affiliation{Centro de Física de Materiales (CSIC-UPV/EHU), Donostia-San Sebastian, Spain}
\affiliation{Department of Physics, University of the Basque Country (UPV/EHU), Bilbao, Spain}
\author{Ion Errea}\affiliation{Centro de Física de Materiales (CSIC-UPV/EHU), Donostia-San Sebastian, Spain}
\affiliation{Donostia International Physics Center, Donostia-San Sebastian, Spain}
\affiliation{Fisika Aplikatua Saila, Gipuzkoako Ingeniaritza Eskola, University of the Basque Country (UPV/EHU), Donostia-San Sebastian, Spain}
\author{Dong Chen}\affiliation{Max Planck Institute for Chemical Physics of Solids, Dresden, Germany}\affiliation{College of Physics, Qingdao University, Qingdao, China}
\author{Maia G. Vergniory}\affiliation{Donostia International Physics Center, Donostia-San Sebastian, Spain}
\affiliation{Max Planck Institute for Chemical Physics of Solids, Dresden, Germany}
\author{Claudia Felser}\affiliation{Max Planck Institute for Chemical Physics of Solids, Dresden, Germany}
\author{Mark H. Fischer${}^{\dagger}$}\affiliation{Department of Physics, University of Zürich, Zürich, Switzerland}
\author{Titus Neupert${}^{\dagger}$}\affiliation{Department of Physics, University of Zürich, Zürich, Switzerland}
\author{Philip J. W. Moll${}^{\dagger}$}\affiliation{Max Planck Institute for the Structure and Dynamics of Matter, Hamburg, Germany}
\date{\today}

\maketitle
\beginsupplement
	
\subsection*{Crystal synthesis and characterization}

 CsV$_3$Sb$_5$ crystallizes in the hexagonal structure with P6/mmm space group. It contains layers of kagome planes formed by the V-atoms. Following a self-flux procedure described in Ref.\cite{ortiz2020cs}, we obtained plate-like single crystals with typical dimensions of 2 × 2 × 0.04 mm$^3$. \\
 The micro-devices S1, S2 and S3 are fabricated using focused-ion-beam (FIB) technique(Fig. S1). Firstly, three lamellae are obtained from the same piece of bulk sample, the plane of the lamella is aligned accurately to the kagome plane with less than $\pm~0.5~$deg misalignment. For device S1, the lamella is transferred in-situ by a micro-manipulator and welded to a gold-coated SiN$_x$ membrane chip via Pt-deposition. Soft springs are later fabricated in order to relax the thermal contraction strain. Meanwhile for both S2 and S3 the lamellae are transferred ex-situ and glued down to a sapphire substrate with pre-patterned gold electrodes. After the fabrication of device S3, a small glue droplet is added on top of the device in order to reduce and homogenize the uniaxial strain observed in S2.
 
Resistance measurements were performed using the resistivity measurement option in a Dynacool PPMS system with a maximal magnetic field of 9~T and base temperature of 2~K. The configuration of in-plane resistance measurements were illustrated in Fig. 1. For each current configuration, the voltage were measured along both side of the devices to ensure the current homogeneity. For all measurements a low AC current of 100 $\mu$A is used to avoid Joule heating. All temperature-dependence measurements were performed with a low sweeping rate of 1~K/min. For all devices, the resistance varies from 0.2 $\Omega$ to 0.2 $\mu\Omega$ with a high residual resistivity ratio above 300 (Fig. S2). To exclude the role of in-plane fields and misalignment of the magnetic field, full two-axis rotations were achieved via a standard PPMS rotator probe: one angle $\theta$ is set by a stepper motor while the angle $\phi$ is controlled by manual positioning outside of the cryostat. The exact angle is determined using the image taken under an optical microscope. 

\subsection*{Detailed analysis of strain effect}
Each device displays a strain profile due to the different geometry. Device S1 is a membrane-based device with micro-springs fabricated by FIB. These soft springs strongly compensate the thermal contraction strain which therefore results in a nearly strain-free situation (Fig. S3). On the contrary, Device S2 is directly attached to the sapphire substrate with glue. The mismatch of thermal contraction coefficient between CsV$_3$Sb$_5$ and sapphire results in a significant uniaxial strain\cite{Maarten,Maja}, as demonstrated by the finite element simulation of strain distribution profile using COMSOL multiphysics software (Fig. S4). The uniaxial strain strongly influences the transport anisotropy of the device. Device S3 features the same setup with S2 yet a glue droplet is applied onto the top after the fabrication process. The larger thermal contraction coefficient of the glue results in a partially compensated and homogenized strain pattern. As a consequence, only the device S2 displays clear anisotropy after the charge order is formed while both S1 and S3 displays no anisotropy at zero field down to low temperature.

\subsection*{Temperature dependence of transport anisotropy and magnetoresistance}

The temperature dependence of the in-plane anisotropy under various magnetic fields (Fig. S5) demonstrates its strengthening in field. Not only the absolute value increases, also the onset at which the anisotropy becomes visible  increases from 25 K at $B$ = 1 T to 70 K at $B$ = 9 T. Note that this temperature dependence is slightly different from the uni-axial strain case as the anisotropy never becomes fully negative. 

The temperature dependence of the magnetoresistance (MR), [R(B)/R(0)], is consistent with previous report as it becomes significant below $T_{CDW}$ and reaches around 10 at base temperature (Fig. S6)\cite{huang2022mixed}. This indicates the Fermi surface reconstruction due to charge ordering, particularly the appearance of 3D Fermi pockets as a signature of enhanced interlayer coupling. 

This trend continues when both magnetic field and strains are applied simultaneously. The transport anisotropy of device S2 increases when an external magnetic field of $B$ = 1 T is applied (Fig. S7).

\subsection*{Angular dependence of magnetoresistance}
In magneto-transport experiments, field misalignment is a common extrinsic source of lowered apparent symmetry. In our case, to examine the effect of field misalignment away from the c-direction on the anisotropic transport we report, an extensive angular dependence of MR measurements have been performed with not one but two rotational axis in order to cover all possible field directions in 3D space (Fig. S8). The rotation from out-of-plane to in-plane direction with the angle $\theta$ is controlled by the rotator motor on the PPMS horizontal rotator with a 0.5 deg/step resolution. Meanwhile, the in-plane rotation is done by manually altering the position of the device on the rotator puck. For each measurement a high-resolution image is taken via the optical microscope, and the angle $\phi$ is determined digitally using the alignment marker on the image. With external field $B$ = 9 T, except for the clear dip around $\theta~\approx$ 180 deg, note that at smaller $\phi$, the MR measured with I//a$_2$ is larger than the MR for other two current directions. The difference is gradually suppressed and becomes invisible for $\phi$ larger than 46 degrees. 

This behavior is clearly in close relation with the longitudinal MR component sets by the angle between magnetic field and current direction, which explains the variation of MR difference with in-plane rotation. The dip at $\theta~\approx$ 180 deg, however, is insensitive to in-plane rotation, as it can be consistently observed in all $\phi$ angles. This demonstrates an intrinsically lower magnetoresistance for I//a$_2$ with $B$//c, which is insensitive to misalignment between the sample and magnetic field. This point can be further elaborated by the results measured at lower magnetic field ($B$ = 1 and 4 T) as a clear difference between I//a$_2$ and other directions can be clearly observed over a wide range of angle $\theta$. This is majorly due to the reduced effect of longitudinal MR component and smaller quantum oscillation amplitude at lower magnetic field. With these evidences, we conclude that the in-plane anisotropy we observed is an intrinsic property of CsV$_3$Sb$_5$ which is robust against extrinsic perturbations such as misalignment.

\subsection*{Field dependence of magnetoresistance}
The magnetoresistance has been measured along all three different current directions. The resistance for I//a$_2$ displays a clear difference compared to the other two curves, which the raw data from which the anisotropy displayed in Fig.3a was extracted.  Meanwhile all curves display a shoulder-like feature at $B$ $\approx$ 1.6 T (Fig. S9), consistent with previous reports\cite{huang2022mixed}. This feature corresponds well with a plateau in the field dependence of transport anisotropy (Fig. 3a). 


\subsection*{Ginzburg-Landau theory}
We consider the order parameters $\Delta_i$ and $\Delta_i'$ as defined in Fig.~2a, which have the transformation properties listed in Tab.~\ref{tab:trnfs}.

\begin{table}[h]
\centering
\caption{Transformation properties of the order parameters $\Delta_i$ and $\Delta_i'$, where $t_i$ are translations by a lattice vector, $C_n$ are rotations and $\sigma_{vi}$ are three mirror planes.}
\begin{tabular}{p{1.7cm} p{1.3cm} p{1.3cm} p{1.3cm} p{1.3cm} p{1.3cm} p{1.3cm} p{1.3cm} p{1.3cm} p{1.3cm}}  
\toprule
& $t_1$ & $t_2$ & $t_3$ & $C_6$ & $C_3$ & $C_2$ & $\sigma_{v1}$ &  $\sigma_{v2}$ &  $\sigma_{v3}$  \\ \hline
\hline
$\Delta_1$&$\Delta_1$&$-\Delta_1$&$-\Delta_1$&$\Delta_3$&$\Delta_2$&$\Delta_1$&$\Delta_1$&$\Delta_3$&$\Delta_2$\\
$\Delta_2$&$-\Delta_2$&$\Delta_2$&$-\Delta_2$&$\Delta_1$&$\Delta_3$&$\Delta_2$&$\Delta_3$&$\Delta_2$&$\Delta_1$\\
$\Delta_3$&$-\Delta_3$&$-\Delta_3$&$\Delta_3$&$\Delta_2$&$\Delta_1$&$\Delta_3$&$\Delta_2$&$\Delta_1$&$\Delta_3$\\\hline
$\Delta_1'$&$\Delta_1'$&$-\Delta_1'$&$-\Delta_1'$&$\Delta_3'$&$\Delta_2'$&$\Delta_1'$&$-\Delta_1'$&$-\Delta_3'$&$-\Delta_2'$\\
$\Delta_2'$&$-\Delta_2'$&$\Delta_2'$&$-\Delta_2'$&$\Delta_1'$&$\Delta_3'$&$\Delta_2'$&$-\Delta_3'$&$-\Delta_2'$&$-\Delta_1'$\\
$\Delta_3'$&$-\Delta_3'$&$-\Delta_3'$&$\Delta_3'$&$\Delta_2'$&$\Delta_1'$&$\Delta_3'$&$-\Delta_2'$&$-\Delta_1'$&$-\Delta_3'$
\label{tab:trnfs}
\end{tabular}
\end{table}

The general (Ginzburg-Landau) free energy including coupling to an out-of-plane magnetic field $B_c$ and strain in the form of a strain tensor $\epsilon_{ij}$ is \cite{Denner,Grandi,Tazai}
\begin{equation}
\begin{split}
    \mathcal{F}[\vec{\Delta}, \vec{\Delta}']=&\alpha(T-T_{\rm c})(\Delta_1^2+\Delta_2^2+\Delta_3^2)+\alpha'(T-T_{\rm c}')(\Delta_1'^2+\Delta_2'^2+\Delta_3'^2) \\ &+\beta_1\Delta_1\Delta_2\Delta_3+\beta_2(\Delta_1\Delta_2'\Delta_3'+\Delta_1'\Delta_2\Delta_3'+\Delta_1'\Delta_2'\Delta_3)\\
    &+\mu_1B(\Delta_1\Delta_1'+\Delta_2\Delta_2'+\Delta_3\Delta_3')\\
    &+\mu_2[(\epsilon_{xx}-\epsilon_{yy})(\Delta_1^2-\Delta_2^2/2-\Delta_3^2/2)+\epsilon_{xy}\sqrt{3}(\Delta_2^2-\Delta_3^2)]\\
    &+\mu_3[(\epsilon_{xx}-\epsilon_{yy})(\Delta_1'^2-\Delta_2'^2/2-\Delta_3'^2/2)+\epsilon_{xy}\sqrt{3}(\Delta_2'^2-\Delta_3'^2)],
\end{split}
\end{equation}
where we have dropped fourth-order terms for simplicity. For the calculation of the full phase diagram, these are chosen in such a way that the perturbation-free case yields an isotropic solution, where we define the anisotropy as $\sum_{i\neq j}(\Delta_i-\Delta_j)^2+\sum_{i\neq j}(\Delta_i'-\Delta_j')^2$. Let us note three important aspects of this free energy: (1) The anisotropy can be induced either through the third-order terms proportional to $\beta_1$ and $\beta_2$ (the latter of which is only active when both order parameters are present) or due to the explicit strain coupling proportional to $\mu_2$ and $\mu_3$. (2) $\vec{\Delta}'$ always induces a subsidiary $\vec{\Delta}$ due to the third-order term proportional to $\beta_2$, but not the other way around. (3) in the presence of a magnetic field, $\vec{\Delta}$ and $\vec{\Delta}'$ are always coupled linearly and therefore appear together. 

Taking (1) and (2) together explains the (perturbation-free) phase diagram in Fig.~2b, which shows the different scenarios as a function of $T_{\rm c}-T_{\rm c}'$. For $T_{\rm c}'<T_{\rm c}$, $\vec{\Delta}$ and $\vec{\Delta}'$ always appear together due to the $\beta_2$ term, such that time-reversal symmetry is broken. $\vec{\Delta}$ is much smaller than $\vec{\Delta}'$, since it is induced by the latter. For lower temperatures, or when $T_{\rm c}'\sim T_{\rm c}$, $\vec{\Delta}$ increases and we cross over to a regime where both order parameters have a comparable magnitude. Finally, when $T_{\rm c}>T_{\rm c}'$ (and in the absence of strain and magnetic field), $\vec{\Delta}$ but not $\vec{\Delta}'$ is non-zero at the charge-ordering temperature and time-reversal symmetry is preserved. 

The inset of Fig. 4 is obtained by numerically minimizing the free energy for a representative set of the coefficients (Tab.~\ref{tab:coefs}), corresponding to scenario (4) in Fig.~2b and computing the anisotropy as a function of $T$. While there is no anisotropy in the pristine case (without strain or $B_c$), applying a magnetic field induces $\vec{\Delta}'$, which in turn leads to anisotropy in the order parameter due to the $\beta_2$ term. Finally, strain trivially leads to an anisotropy via the $\mu_2$ term. 

\begin{table}
\centering
\caption{Choice of Ginzburg-Landau coefficients used for Fig.~3c and Fig.~4 inset.}
\begin{tabular}{p{2.5cm} p{1.3cm} p{1.3cm} p{1.3cm} p{1.3cm} p{1.3cm} p{1.3cm} p{1.3cm} p{1.3cm} p{1.3cm}}  
\toprule
Parameter & $\alpha$ & $\alpha'$ & $\beta_1$ & $\beta_2$ & $\mu_1$ & $\mu_2$ & $\mu_3$ \\ \hline
Value & 1        & 1         & 1         & $-1$        & 1    & 1       & 1.5    
\label{tab:coefs}
\end{tabular}
\end{table}

In Fig. S10 we show the results for the anisotropy from the GL theory for a different set of GL parameters corresponding to regime (1), where the TRS-breaking order is dominant. This regime is inconsistent with experimental observations, since it shows a suppression of the anisotropy when a magnetic field is applied. 

Not captured by our two-dimensional theory is the possibility that an isotropic response of the pristine system is the result of averaging over many layers with randomly distributed anisotropies (i.e., domains in the stacking direction). However, we consider this possibility unlikely for the following reason. Upon applying the magnetic field, the layers would need to align their anisotropies to yield the experimentally observed macroscopic anisotropy. In contrast, our GL theory would imply that each layer becomes more anisotropic in the magnetic field. The random stacking would then still result in an averaging to an isotropic response.  

\clearpage

\clearpage
\renewcommand{\figurename}{Fig.}
\setcounter{figure}{0} 
\begin{figure}
	\centering
\includegraphics[width = 0.98\linewidth]{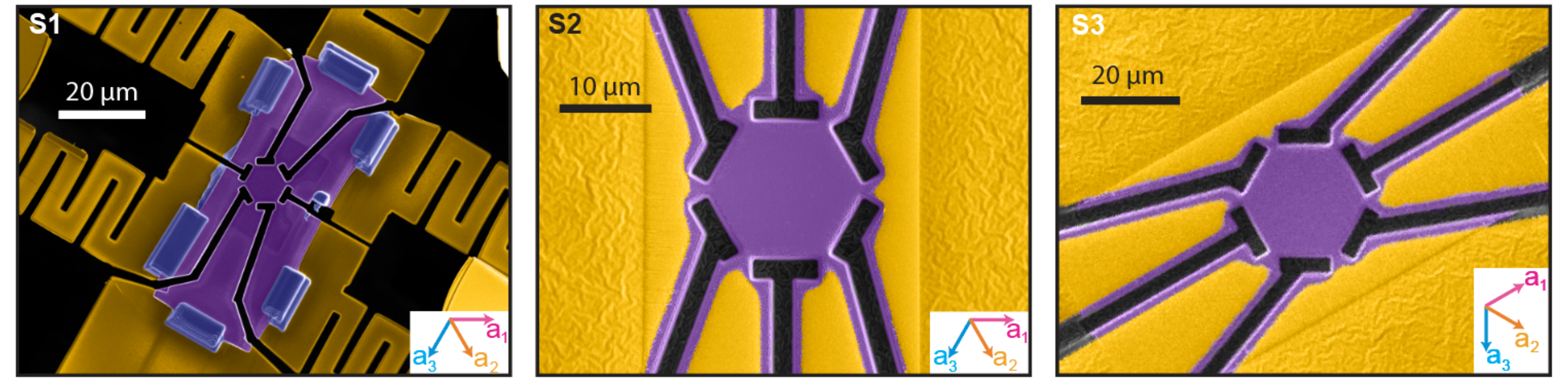}
		\caption{\textbf{Scanning electron microscope (SEM) images of  device S1, S2 and S3.} S1 is a membrane device with reduced differential thermal contraction strain, while S2 is glued down to a sapphire substrate, the coupling between the substrate and sample results in the non-negligible uni-axial strain at low temperature. S3 is a similar device with S2, yet after taking the SEM image, a glue droplet is applied on top of the device to reduce uniaxial strain at low temperature.}
	\label{SEM}
\end{figure}	
\clearpage

\begin{figure}
	\centering
\includegraphics[width = 0.9\linewidth]{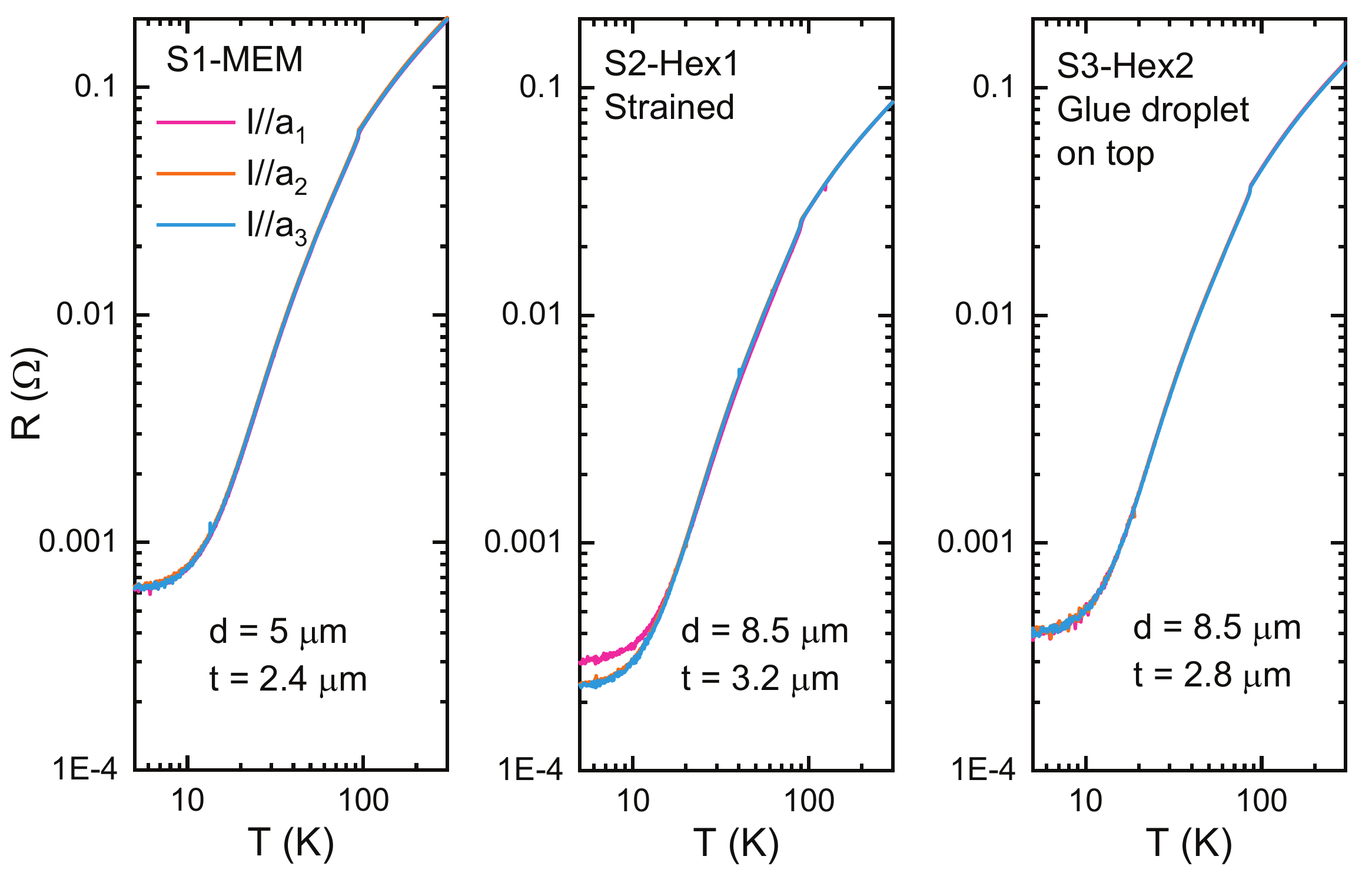}
		\caption{\textbf{Temperature-dependent resistance of all three different devices.} d and t stand for the edge length and thickness of the hexagon respectively. For both S1 and S3, the T-dependence of resistance measured along different axis is nearly identical with each other. This demonstrates the isotropic in-plane transport in these two devices. Meanwhile, for device S2, the resistance measured with current along a$_1$ displays a clear difference with the results measured along other current directions. This difference stands for the electronic anisotropy which increases with decreasing temperature.} 
	\label{basicR}
\end{figure}

\begin{figure}
	\centering
\includegraphics[width = 0.9\linewidth]{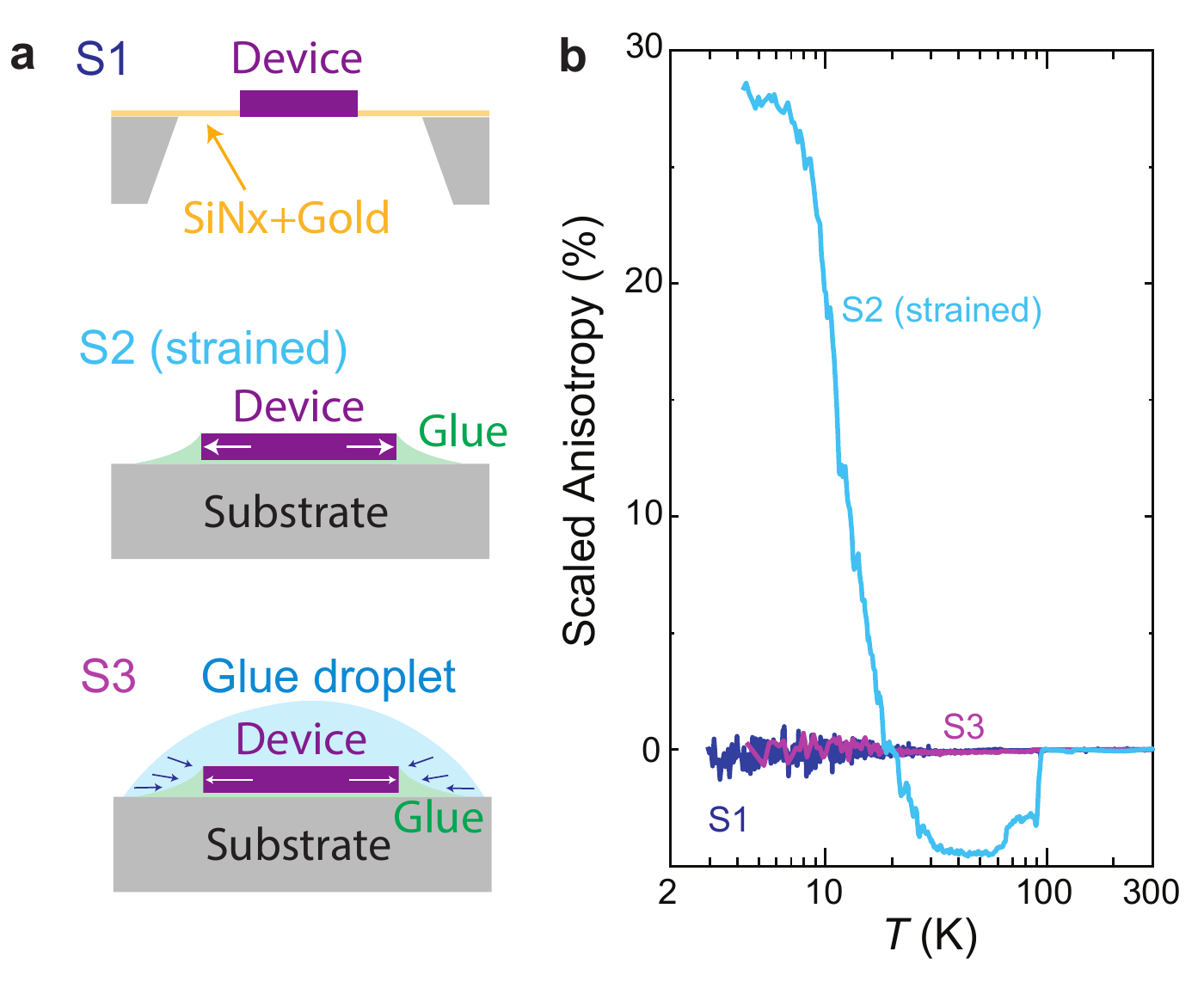}
		\caption{\textbf{Strain analysis and T-dependence of transport anisotropy.} (a) displays the detailed setup of each devices. The uniaxial strain due to differential thermal contraction is suppressed by the soft membrane spring in device S1 and compensation of glue droplet in device S3. (d) Temperature dependence of transport anisotropy of all devices. Both S1 and S3 display negligible anisotropy down to low temperature, while the strained device S2 shows clear anisotropy which appears at the charge-density-wave transition and gets more pronounced at base temperature. } 
	\label{strain}
\end{figure}

\begin{figure}
	\centering
\includegraphics[width = 0.9\linewidth]{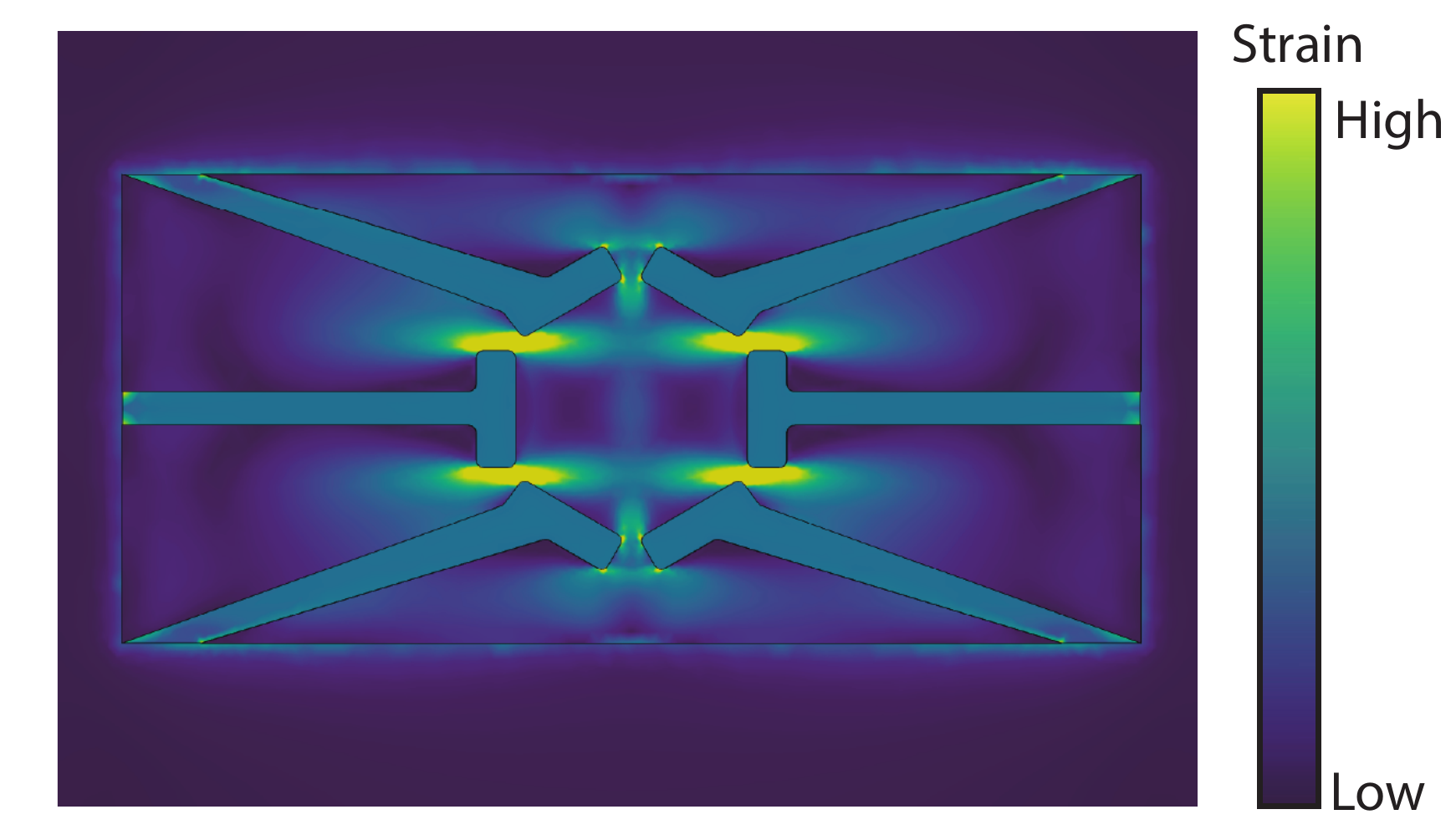}
		\caption{\textbf{COMSOL simulation of uniaxial strain.} Finite element simulation of strain distribution of device S2 based on COMSOL multiphysics. The results display a dominant uniaxial strain component along the long side of the lamella.} 
	\label{Comsol}
\end{figure}

\begin{figure}
	\centering
\includegraphics[width = 0.9\linewidth]{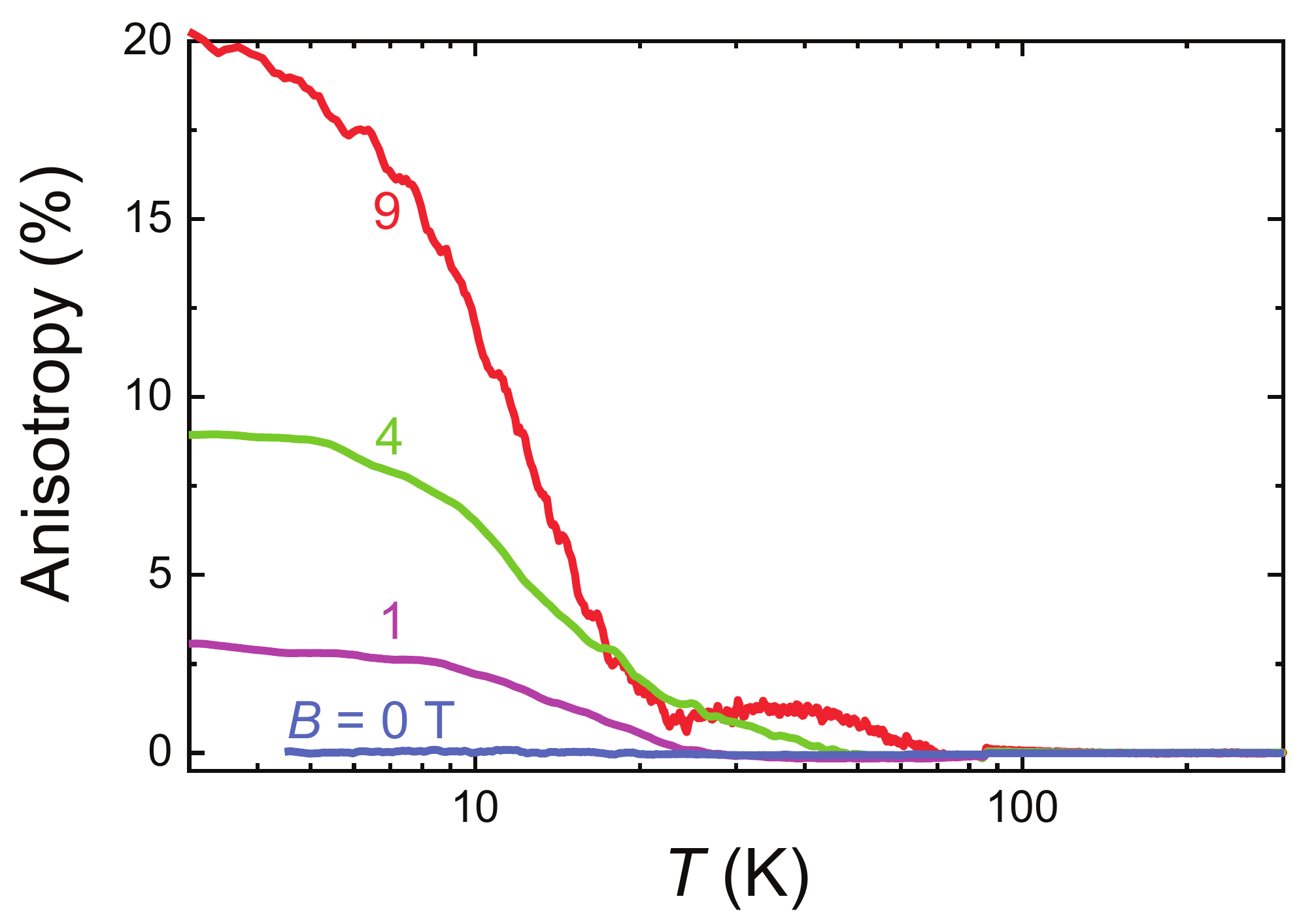}
		\caption{\textbf{Field-induced transport anisotropy of S3.} T-dependent transport anisotropy of S3 measured at different magnetic fields. The low temperature anisotropy increases strongly with increasing magnetic field.}
	\label{B-dep}
\end{figure}

\begin{figure}
	\centering
\includegraphics[width = 0.9\linewidth]{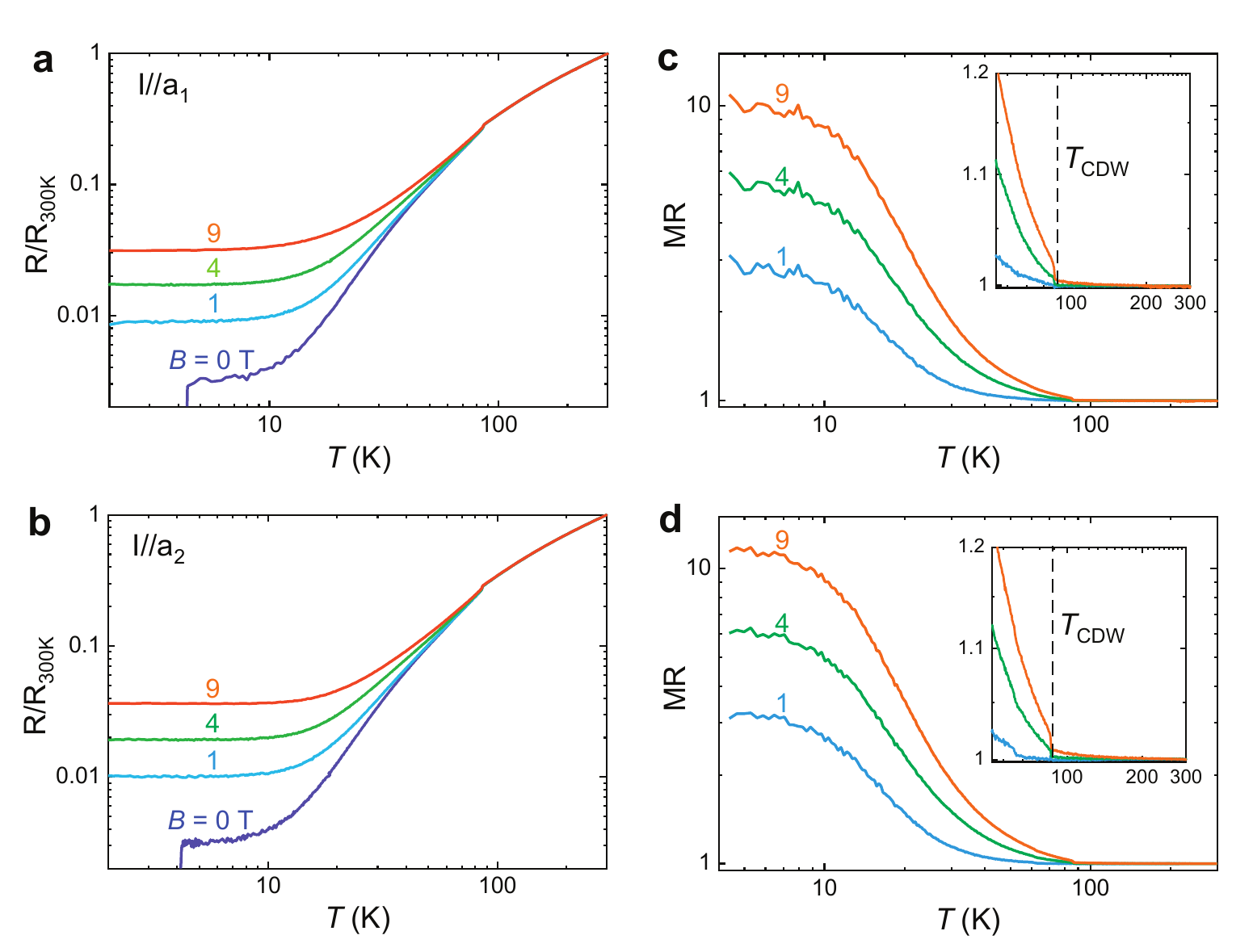}
		\caption{\textbf{T-dependence of magnetoresistance for S3.} (a) and (b) display the T-dependence of resistance measured at various magnetic fields with current along a$_1$ and a$_2$ respectively. The magentoresistance ratio [R(B)/R$_0$] for both current directions are presented in (c) and (d) respectively. It is clear that MR only becomes significant below the charge-ordering temperature.}
	\label{MR}
\end{figure}

\begin{figure}
	\centering
\includegraphics[width = 0.9\linewidth]{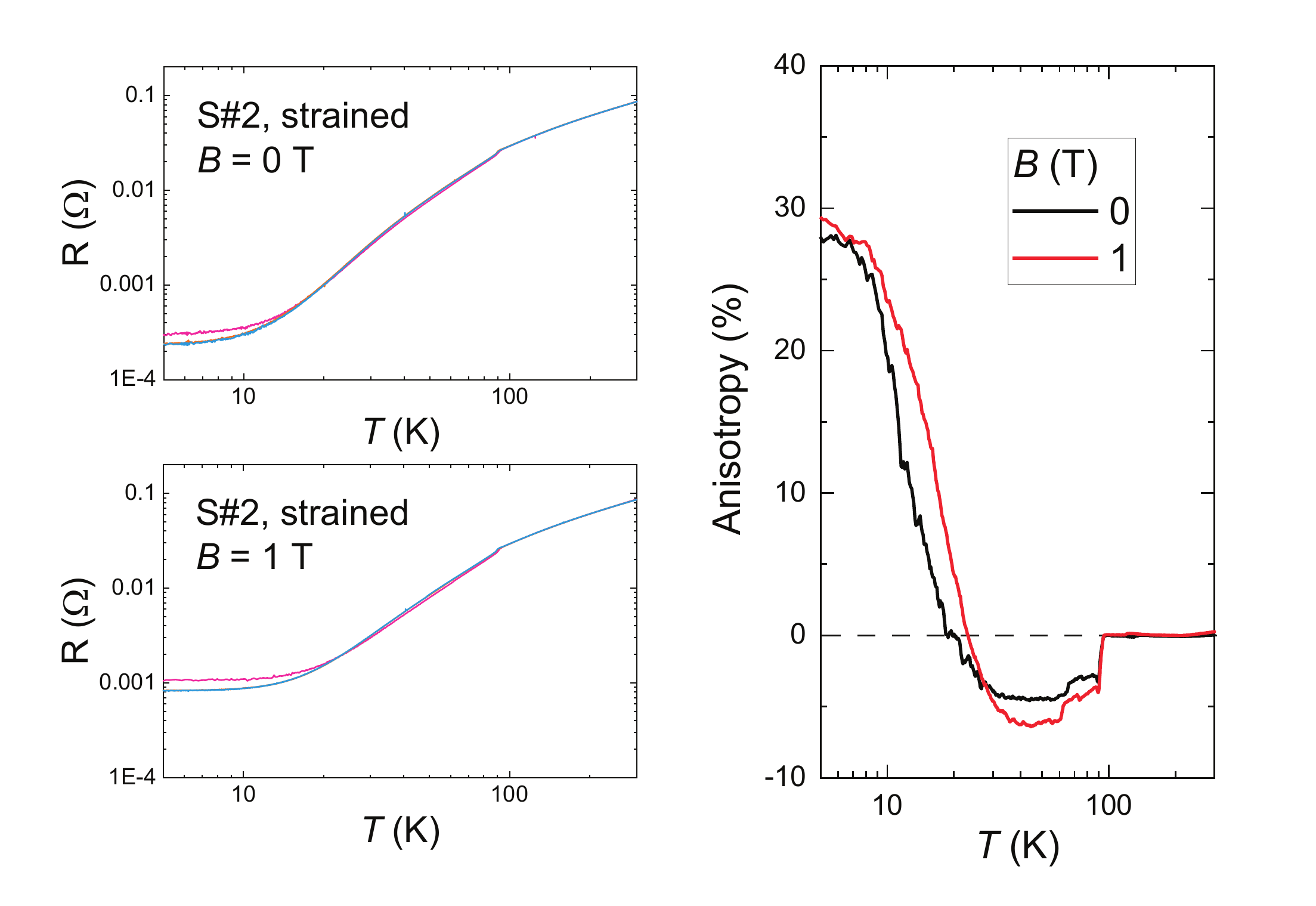}
		\caption{\textbf{T-dependent resistance and in-plane anisotropy of S2 at $B$ = 0 and 1 T.} (a) and (b) present T-dependence of directional resistance measured at zero field and $B$ = 1 T. The anisotropy due to strain is slightly enhanced by magnetic field, as shown in (c). }  
	\label{strain+field}
\end{figure}

\clearpage

\begin{figure}
	\centering
\includegraphics[width = 0.9\linewidth]{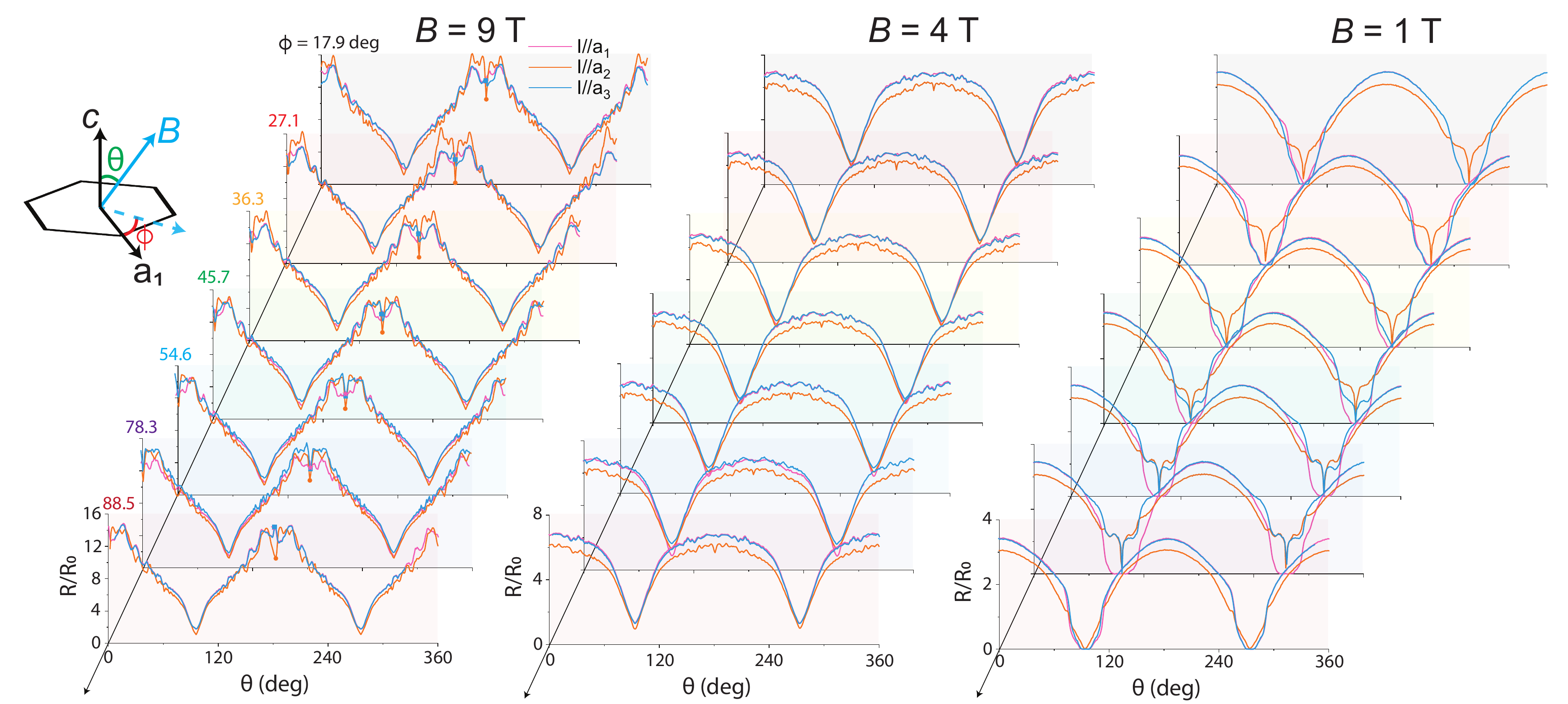}
		\caption{\textbf{Angular dependence of MR (AMR) for S3.} AMR measured with spherical rotation and different magnetic fields. The difference between I//a$_2$ and other two current directions can be clearly seen for all $\phi$, demonstrating the intrinsic nature of such behavior. }
	\label{AMR}
\end{figure}
\clearpage

\begin{figure}
	\centering
\includegraphics[width = 0.5\linewidth]{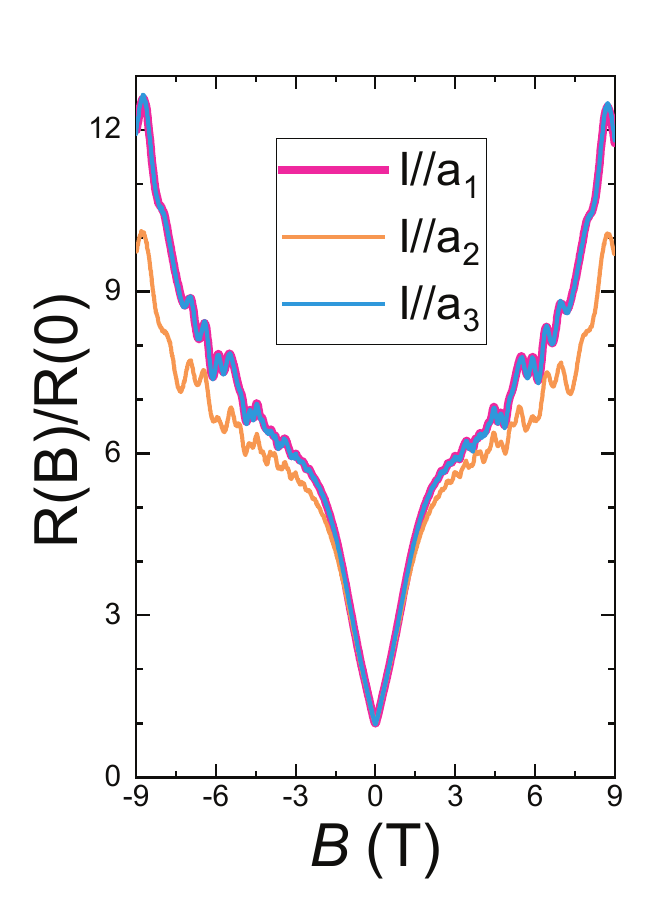}
		\caption{\textbf{Magnetoresistance along different directions.} Field dependence of magnetoresistance measured along all three different current directions at $T$ = 2 K in device S3. The results measured with I//a$_1$ and I//a$_3$ overlaps exactly with each other, yet deviates clearly with I//a$_2$. All magnetoresistance displays a shoulder-like feature at around $B$ $\approx$ 1.6 T. }
\end{figure}

\begin{figure}
	\centering
 \includegraphics[width = 0.5\linewidth]{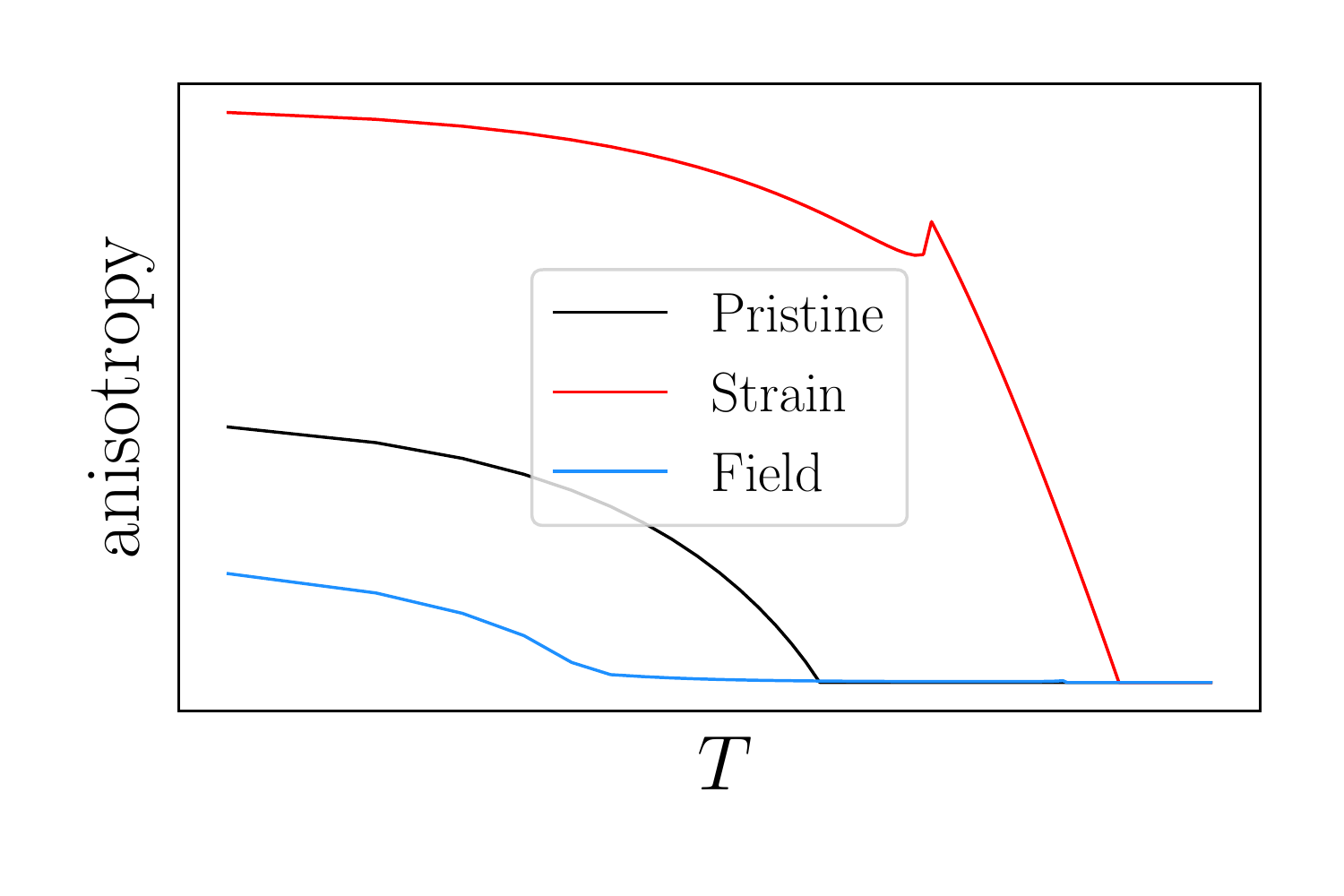}
		\caption{\textbf{Anisotropy from GL theory in regime \textcircled{1}.} The pristine, strained and magnetic field cases all show an anisotropy and in particular the magnetic field suppresses the anisotropy contrary to the experimental observations. The anisotropy in regime \textcircled{1} is therefore not consistent with experiments.}
\end{figure}